\documentclass[usenatbib]{mn2e}

\usepackage{subfig}
\usepackage{graphicx}
\title[Mid-IR Quasar Templates]{Characterizing Quasars in the Mid-infrared: High Signal-to-Noise Spectral Templates}

\author[Allison. R. Hill et al.]
       {Allison.~R.~Hill,$^{1}$\thanks{E-mail: ahill49@gmail.com}
       S.~C.~Gallagher,$^{1}$ R.~P.~Deo,$^{1}$ E.~Peeters$^{1,2}$ and 
       Gordon.~T.~Richards$^{3,4}$\\
       $^{1}$The University of Western Ontario, 1151 Richmond Street, London, ON, N6A3K7, Canada\\
       $^{2}$SETI Institute, 189 Bernardo Avenue, Suite 100, Mountain View, CA 94043, USA\\
       $^{3}$Department of Physics, 3141 Chestnut Street, Drexel University, Philadelphia, PA 19104, USA\\
       $^{4}$Max Planck Institut f\"{u}r Astronomie, K\"{o}nigstuhl 17, Heidelberg, Germany 69117}

\begin{document}

\date{Accepted (blank). Received (blank)}
\pagerange{\pageref{firstpage}---\pageref{lastpage}} \pubyear{2013}

\maketitle

\label{firstpage}

\begin{abstract}

  Mid-infrared (MIR) quasar spectra exhibit a suite of emission
  features including high ionization coronal lines from the narrow
  line region illuminated by the ionizing continuum, broad dust bumps
  from silicates and graphites, and polycyclic aromatic hydrocarbon
  (PAH) features from star formation in the host galaxy. However, in
  \textit{Spitzer} Infared Spectrograph (IRS) data, few features are
  detected in most individual spectra because of typically low
  signal-to-noise ratios (S/N).  By generating spectral composites
  from over 180 IRS observations of Sloan Digital Sky Survey
  broad-line quasars, we boost the S/N and reveal features in the
  complex spectra that are otherwise lost in the noise.  In addition
  to an overall composite, we generate composites in three different
  luminosity bins that span the range of 5.6~\micron\/ luminosities of
  10$^{40}$--10$^{46}$ (erg~s$^{-1}$).  We detect the high-ionization,
  forbidden emission lines of [SIV], [OIV], and
  [NeV]~$\lambda14~\micron$ in all templates and PAH features in all
  but the most luminous template.  Ratios of lines with a range of
  ionization potentials show no evidence for a strong difference in
  the shape of the 41--97~eV ionizing continuum over this range
  of luminosities.  The scaling of the emission-line luminosities as a
  function of continuum luminosity is consistent with what is
  expected, and shows no indication of a ``disappearing narrow-line
  region.'' The broad 10 and 18~\micron\/ silicate features in
  emission increase in strength with increasing luminosity, and a
  broad 3--5~\micron\/ black body consistent with graphite emission at
  750~K is evident in the highest luminosity template.  We find that
  the intrinsic quasar continua for all luminosity templates are
  consistent; apparent differences arise primarily from host galaxy
  contamination most evident at low luminosity.

\end{abstract}


\section{Introduction}

Quasars -- growing supermassive black holes in the centers of massive
galaxies -- are the subset of active galactic nuclei (AGNs) that
constitute the most luminous objects in the universe.  They radiate
substantial power across much of the electromagnetic spectrum, with
the source of radiation in each frequency regime originating from a
different location with respect to the supermassive black hole. The
shape of a quasar's spectral energy distribution (SED) can reveal much
about the structure of the black hole-accretion disk system.  Prior to
the sensitivity of {\it Spitzer} in the mid-infrared (MIR), quasar
SEDs covering this wavelength regime could only be constructed for
small samples of the brightest objects. \citet{elv94} used
multiwavelength photometry of 47 objects to generate a mean quasar SED
which covered much of the electromagnetic spectrum (from radio to hard
X-rays). The average SED showed some characteristic continuum features
including the UV and MIR `bumps'. The source of the UV bump is best
explained as optically thick continuum emission from the accretion
disk itself \citep[e.g.,][]{mal82}. The MIR bump is the thermal emission from
dust grains heated by the accretion disk continuum.  For dust to
survive, the source of this emission must be at much greater distances
($\geq 1~\textrm{pc}$) from the black hole than the accretion disk.

Quasars and AGN can be classified as Type 1 or Type 2 based on the
absence (Type 2) or presence (Type 1) of broad emission lines in their
optical spectra. The most persistent model to explain this difference
in recent decades includes a dusty torus, that when viewed close to
edge-on obscures the broad-line region close to the central black hole
\citep[e.g.,][]{ant93, urr95}. This unified model implies no inherent
difference between Type 1 and Type 2 objects; the effect is merely
geometric. (The dynamical stability of the torus is an issue which is
not addressed in the aforementioned unification picture.) An
alternative view to simple unification is that the obscuring `torus'
is actually an outflow of dusty material from a disk wind
\citep[e.g.,][]{kon94,eli06,kea12}. In this picture, the differences
between Type 1 and Type 2 objects may not be purely geometric (e.g.,
if the dust distribution is patchy) and/or may have a dependence on
evolution or accretion rate (or both).  Dusty torus or not, there
exists a distance from the black hole-accretion disk system where
temperatures are low enough that dust can survive; this distance from
the central engine is the dust sublimation radius ($R_{sub}$). Beyond
this point, the dust may be heated by and thus reprocess emission from
the accretion disk, achieving peak temperatures of 100--1500~K
\citep[e.g.,][]{ant93}. At these temperatures, the peak of the
emission will be in the MIR, thus accounting for the continuum
emission of the MIR bump present in quasar SEDs
\citep[e.g.,][]{elv94}.

The depth and sky coverage of {\it Spitzer} photometry enabled the
construction of multiwavelength quasar SEDs from larger and more
complete samples.  From 259 objects, \citet{ric06b} created average
SEDs, binning the data to search for trends with parameters such as
luminosity, radio-loudness, and optical color. They found the SEDs to be very
similar with the exception of a near-infrared excess at $\sim4$
\micron\ in the most luminous quasars (a difference also noted by
\citealt{hat05} in a smaller sample of 35 objects). This excess flux
detected in the near-IR spectra of some AGN has been noted in the
literature for decades \citep{ede86}. More recently, \citet{gal07}
observed the same bump in a sample of 234 quasars as a steepening of
the spectral index in the $1-8~\micron$ range, explained as a thermal
black body of hot graphite dust. Graphites have a higher dust
sublimation temperature ($\sim 1500~\textrm{K}$) than the silicates
which are the source of the 10 and 18~$\textrm{\micron}$ features, and
therefore have a smaller $R_{sub}$ and are located closer to the
central engine than silicates. While trying to model individual
luminous quasar IRS spectra, \citet{deo11} found the fits often
required a blackbody in the $2-8~\textrm{\micron}$ range with a
temperature of $\sim 1200~\textrm{K}$, further supporting the idea of
a hot dust component.

Cooler dust emits thermal continuum at longer wavelengths overlaid
with broad features roughly centred at 10 and 18 \micron\ from
vibrational modes of silicate grains.  Using {\it Spitzer}
observations, \citet{hao05} detected both the 10 and 18 \micron\/
features in emission in a sample of Type 1 PG quasars. This result
seemed to support the picture proposed by \citet{ant93} with the
silicates in emission caused by a viewing angle of the `torus' which
is face-on in Type~1 objects, where the inner wall of the torus is
illuminated (as opposed to an edge-on view where the column density is
much higher and the silicates are seen in absorption).
  
In addition to the thermal continuum emission in the MIR,
high-ionization emission lines from the forbidden transitions of ions
such as [SIV], [OIV], and [NeV] were expected \citep{spi92} from the
ionization and excitation of the low density host galaxy ISM by the UV
through soft X-ray continuum of the accretion disk that is not blocked by
the `torus'.  Many of the high ionization lines predicted by
\citet{spi92} were observed in {\em ISO} studies \citep{gen98}, and
early spectra of MIR bright quasars from {\em Spitzer}
\citep[e.g,][]{wee05}.  While there have been many studies of emission
line properties from the narrow lines in nearby AGNs
\citep[e.g.,][]{ker09,per10,das11}, the studies of archetypal luminous
quasars at $z\sim$1--2 have largely been limited to continuum studies;
as an example, \citet{deo11} detected no forbidden lines in their 25
IRS spectra due to the relatively low $S/N$ from these fainter
objects.

It was not until recently that enough data have been gathered to
perform a MIR spectroscopic statistical study. Because of the poor $S/N$
in many spectra, it became almost a necessity to study features in
quasars through spectral composites, as one of the only ways to
reveal faint, ubiquitous features that would otherwise be lost in the
noise. \citet{hao07} took a sample of 196 AGNs and ultra-luminous
infrared galaxies (ULIRGs) from the {\it Spitzer} archive; of that
subset, 24 were quasars, while the rest of the sample contained
Seyferts (1 and 2) and ULIRGs. Average spectra were generated for each
category of object, exhibiting the 10 and 18~\micron\/ features in
emission, along with weak detections of several MIR forbidden lines
and PAH features. Compared to Seyferts and ULIRGs, the detected lines
in the average luminous quasar spectrum (forbidden and PAH features
alike) had smaller equivalent widths (EW).\footnote{This is reminiscent of the
  Baldwin Effect \citep{bal77} where the EW of the optical broad line
  CIV was shown to be anti-correlated with the optical continuum
  luminosity.} The quasar continuum also increases to longer
wavelengths more slowly (in units of $L_{\nu}$ vs. $\lambda$) than the
lower luminosity classes of objects --- i.e., quasar spectra are
`flatter' (i.e., bluer) than Seyferts or ULIRGS. This is confirmed by
\citet{sar11}, who compared the IR spectra of dusty starbursts and AGN
and found ``emission AGN'' (in reference to the 10 \micron\/ feature)
have flatter spectra than ``absorption AGN''.

Quasars have been studied extensively in the optical/UV, as unobscured
quasars are readily studied from the ground in this regime.  Only
space-based facilities such as {\em Spitzer} have the low background
that allow them the sensitivity to detect similar objects in the MIR.
As a result, the optical has the advantage of much larger sample sizes
provided by the Sloan Digital Sky Survey (SDSS) quasar catalog
\citep{sch10}, which contains over a hundred thousand
quasars. \citet{van01} used a sample of over 4000 Type 1 objects to
generate spectral composites covering the rest-frame optical through
UV. Their work determined canonical values for optical and UV line
equivalent widths and spectral indices. We extend this effort into the
MIR for type 1 quasars with a sample size of 184 objects -- thus far
unprecedented in the literature for this wavelength regime.

For this study, the assumed cosmology is
${H}_{0}=70~\textrm{km}~\textrm{s}^{-1}~\textrm{Mpc}^{-1}$,
$\Omega_{\rm M}=0.3$, and $\Omega_{\Lambda}=0.7$.

\section{Sample Selection}
\label{sec:sample}

\begin{figure}
  \centering
  \includegraphics[width=\linewidth]{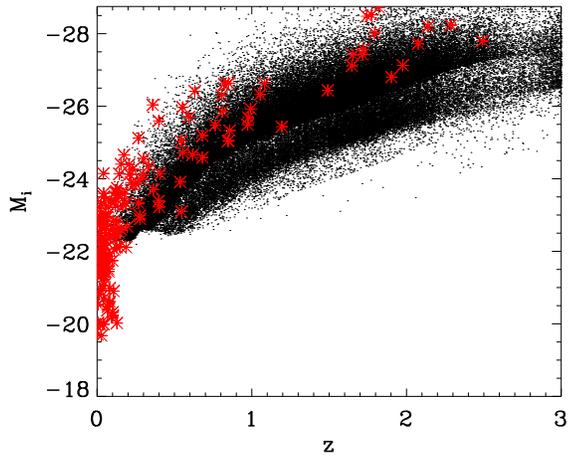}
  \caption{Absolute, $K$-corrected (Table 4; \citealt{ric06a})
    $i$-band magnitudes from \citet{she11} versus redshift. The black
    points correspond to the SDSS quasar catalog (\citealt{she11},
    ${M}_{i}<22$). The red asterisks indicate the 184 objects of our
    sample, with the ${M}_{i}$ values taken from the SDSS DR7
      photometric catalogue \citep{aba09}.}
  \label{fig:diagnostic}
\end{figure}

We coordinate cross-matched the SDSS quasar catalogue
\citep{sch10} with the {\it Spitzer} \citep{wer04} Infrared
Spectrograph \citep[IRS;][]{hou04} archive, using a match radius
  of 2 arcseconds, yielding 184 low resolution (${\it R}\sim60-130$)
MIR spectra (see Table~\ref{tab:data}).  We selected all data which
had short-low (SL; $5.1-14.3~\micron$) and/or long-low (LL;
$13.9-39.9~\micron$) modules (although not all objects necessarily had
both modules), opting to keep only the low resolution data. Although
our data span a wide range of luminosities
($\log(L_{5.6\micron\/})=41.1-46.1~[\textrm{erg/s}]$) and redshifts
($z=0.001-2.49$), most of our objects occupy the low luminosity
(median of $43.9~[\textrm{erg/s}]$ in log units), low redshift (median
of 0.081) regime of the parameter space
(Figure~\ref{fig:diagnostic}). The low luminosity nature of our
objects limits access to various parameters (e.g., virial black hole
masses, and UV line luminosities) tabulated in quasar catalogs such as
\citet{hew10} and \citet{she11}. For example, only 70 of our objects
are in the \citet{she11} catalogue.
 
\begin{figure}
	\centering
	\includegraphics[width=\linewidth]{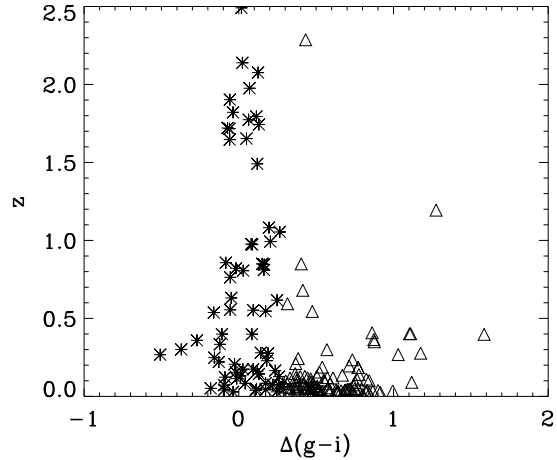}
	\caption{Redshift versus the relative $g-i$ colour of the objects in our sample. Open triangles indicate quasars which are defined as `red', in that $\Delta(g-i)>0.3$. The objects with $\Delta(g-i)<0.3$ are defined as `normal' (following the convention of \citealt{ric03}) and are represented as asterisks. The total number of objects classified as `red' is 116.}
	\label{fig:reddened}
\end{figure}

A further possible bias in our sample is that 116 quasars
have redder UV/optical colors compared to the sample of \citet{ric03},
i.e., they are characterized by a more negative spectral index
$\alpha$ in $f(\nu)\propto\nu^\alpha$. Following the example of
\citet{ric03}, we calculated the relative {\it g-i} colours,
$\Delta(g-i)$ \citep{ric01}, of the quasars in our sample, and plotted
them against their redshifts (Figure~\ref{fig:reddened}). The relative
colour accounts for the effects of redshift as prominent emission
lines shift in and out of the SDSS bands, by determining the
underlying continuum color. The median colors of quasars at the
redshift of each quasar are subtracted from the measured colour of
each object. Adopting the same cutoff as \citet{ric03} we define `red'
quasars as those with $\Delta(g-i)>0.3$. Using this definition, we
find that the majority of our objects are classified as `red'
(Figure~\ref{fig:reddened}). It is not clear if this `reddening' is
intrinsic to the quasar, or is a result of contamination from the host
galaxy in our low redshift sources.  Spectral decomposition would be
required to comment further on the nature of the dust reddening and/or host
galaxy contamination in our low-luminosity, low-redshift objects.

\begin{table*}
  \begin{minipage}{\linewidth}
    \caption{{\it Spitzer} Object Summary (full table available in electronic format)}
    \begin{tabular}{@{}lccccccccc}
      \hline\hline
      & \multicolumn{3}{c}{SDSS Identifiers} & \multicolumn{2}{c}{{\it Spitzer} Identifiers} & & & & \\
      Object~Name\footnote{Objects are sorted in descending order of $\log(L_{5.6\micron})$} & plate & fid & mjd & aorkey & program~ID & z\footnote{Redshifts taken from \citet{hew10}} & $\log(L_{5.6\micron})$ & $M_{i}$\footnote{Absolute i-band magnitude} & Template\footnote{Number denotes which luminosity bin object resides in (where 1 is the most luminous group, 2 is the intermediate group and 3 is the least luminous group)} \\
      & & & & & & & [erg/s] & &\\
      \hline
      FTM1113+1244 & 1604 & 566 & 53078 & 22391040 & 30119 & 0.680 & 46.095 & -24.6 & 1 \\
      SDSSJ151307.75+605956.9 & 351 & 231 & 51695 & 14757376 & 50588 & 2.287 & 46.052 & -28.2 & 1 \\
      SDSSJ170102.18+612301.0 & 613 & 365 & 52345 & 25977088 & 40936 & 1.821 & 46.013 & -28.7 & 1 \\
      SDSSJ132120.48+574259.4 & 760 & 287 & 52264 & 18027776 & 40936 & 1.794 & 45.902 & -28.0 & 1 \\
      SDSSJ024933.42-083454.4 & 1320 & 376 & 52759 & 25977344 & 3237 & 1.774 & 45.890 & -28.5 & 1 \\
      \multicolumn{10}{c}{:} \\
      \hline
      \label{tab:data}
  \end{tabular}
\end{minipage}
\end{table*}

\section{Data Reduction \& Analysis}

\subsection{Spitzer Spectra}

We obtained the basic calibrated data (BCD) products processed with
the standard {\em Spitzer} IRS pipeline (version S18.7.0) from the
{\em Spitzer} Science Center (SSC) archive. We cleaned the BCD images
using the IRSCLEAN software package (part of the SMART \citealt{hig04}
package) to fix rogue pixels using SSC supplied masks, and a weak
thresholding of the pixel histogram to remove highly discrepant rogue
pixels. We co-added the multiple data collection event (DCE) image
files into one image for each module, spectral order, and ÒnodÓ
position (e.g., SL, first order, first nod position) using the ``fair
co-add'' option in SMART \citep{hig04}. The co-added images from the
opposite order/nod positions (as per setup of observation) were
subtracted from each-other to remove the sky background in each. The
spectra were extracted using the ``optimal extraction option'' within
the SMART package. All the image combining and spectrum extraction
operations were carried out using SMART. We checked our extractions
using automated extractions with the {\em Spitzer} IRS Custom
Extractor (SPICE) program also, and found good matches between
extracted spectra. In the cases where the SL and LL spectra of the
  same object did not have the same average flux values where they
  overlapped in wavelength, we scaled the SL module spectra to the LL
  module spectra as the LL slit widths are larger than the SL slit
  widths (cf. \citealt{bra06}).

\subsection{Template Construction}

\begin{figure}
	\includegraphics[width=0.9\linewidth]{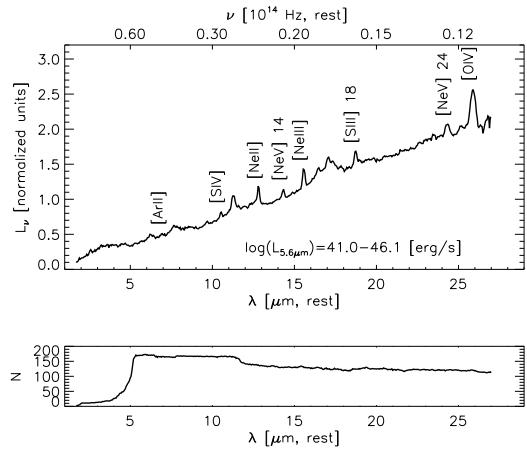}
	\caption{{\it Top panel}: Rest-frame quasar template composed of all objects in the sample. Prominent narrow line emission features are labeled. The 5.6 \micron\ luminosity range is indicated in the lower right. {\it Bottom panel}: Number of objects contributing to the template in each wavelength bin.}
	\label{fig:total_irs}
\end{figure}

To prepare the spectra for averaging, the data need to be shifted to
the rest frame and interpolated to a common wavelength scale. We use a
bin size of $0.05~\micron$. Next, an appropriate continuum point for
normalization was selected to minimize overlap with significant
features such as PAH emission, forbidden lines, and broad silicate
features. \citet{spo07} devised methods to measure the flux densities
of a wide variety of galaxies using the 5.6, 14, and 26~$\micron$
locations as continuum points (chosen because they are relatively
featureless regions of the spectrum).  To highlight the differences in
continuum in the red and blue end of the MIR, we chose 14~\micron\ as
the normalization point as it effectively bisects the wavelength range
considered. Not all of the objects contained flux information at
14~$\micron$, due to either redshift effects, or the absence of
coverage of one or more of the {\it Spitzer} modules. To account for
this, we took advantage of the fact that quasars have similarly shaped
SEDs \citep{ric06b}. We utilized the multi-wavelength SED composite
from \citet{ric06b}, and normalized the data to the 5.6~$\micron$
continuum point and extrapolated along the SED to calculate flux
information at 14~$\micron$. Once shifted and normalized, the average
template spectrum was calculated, with a $3\sigma$-clipped mean
returned for each wavelength bin. The template constructed from all
186 objects can be found in Figure~\ref{fig:total_irs}.

\begin{figure}
	\includegraphics[width=\linewidth]{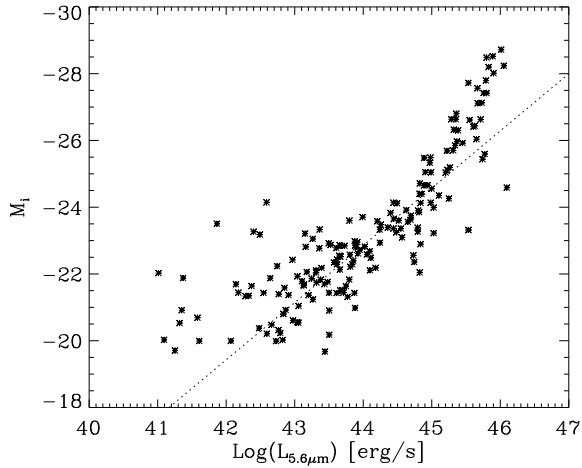}
	\caption{The absolute $i$-band magnitude, $M_{i}$, versus the $L_{5.6\micron\/}$ continuum luminosity, of each quasar. A line of slope 1 is overplotted (dashed line). At low luminosities there is greater scatter, with the values of $M_{i}$ greater than would be expected with a 1:1 relation (assuming both continuum points trace the bolometric luminosity of the quasar). This likely results from the greater sensitivity of $M_{i}$ to  host galaxy contamination compared to $L_{5.6\micron\/}$. At the high luminosity end, objects typically have high redshifts, and $M_{i}$ is calculated from the observed-frame optical photometry with a typical K-correction, and therefore may not be an accurate tracer of the optical continuum luminsosity.}
	\label{fig:lumIR_lumMi}
\end{figure}

To study the effects of luminosity on global trends, the luminosity at
$5.6~\micron$ was calculated for each object. This continuum point was
chosen because it has the most overlap of the rest frame spectra, and
is a better indicator of the quasar luminosity, as contamination from
the host galaxy typically occurs at longer wavelengths in the MIR
\citep{sch06}. We opted for a MIR continuum point over an optical
continuum point because of the large proportion of our objects which
may suffer from host galaxy contamination and/or optical continuum
reddening (Figure~\ref{fig:reddened}). We suspect contamination from
the host to be more noticeable in the optical over the $5.6~\micron$
continuum point, as seen in Figure~\ref{fig:lumIR_lumMi}. 

  At the low luminosity ($L_{\rm 5.6\micron}<10^{43}$ erg~s$^{-1}$)
  end of the distribution in Figure~\ref{fig:lumIR_lumMi}, a large
  fraction of AGNs show an excess of ${M}_{i}$ emission. This is
  likely a result of the significant contribution of host galaxy
  emission in the rest-frame $i$-band.  However, the 5.6~\micron\
  continuum comes from hot (few hundred to a thousand K) dust in the
  immediate vicinity of the quasar; these high temperatures are not
  typical of dust heated by star formation, and thus the host galaxy
  contribution is expected to be negligible. Therefore,
  $\log(L_{5.6\micron\/})$ is a better tracer of the intrinsic
  luminosity of the quasar.  It is less clear what causes an
  excess of $M_{i}$ compared to $\log(L_{5.6\micron\/})$ at the high
  luminosity end in Figure~\ref{fig:lumIR_lumMi}.  However, $M_{i}$ is
  an imperfect measure of optical luminosity, as it is calculated by
  assuming a typical $K$-correction from the observed optical
  photometry.  The extrapolation from the observed photometry to
  $M_{i}$ is larger at higher redshift, and the most luminous objects
  in our sample have (on average) significantly higher redshifts than
  the less luminous AGNs (see Fig.~\ref{fig:diagnostic}).  Using the
  integrated optical and IR photometry of a sample of 234 SDSS quasars
  to determine $L_{\rm opt}$ and $L_{\rm IR}$ from \citet{ric06b},
  \citet{gal07} found that the ratio of the two was consistent with
  being constant as a function of luminosity once dust-reddening
  (which preferentially reduces the optical luminosity) was taken into
  account (see their Fig. 2).  Therefore, we do not consider the
  apparent ``excess'' of optical luminosity as measured by $M_{i}$ to
  indicate a true effect.

\begin{figure}
	\centering
	\includegraphics[width=\linewidth]{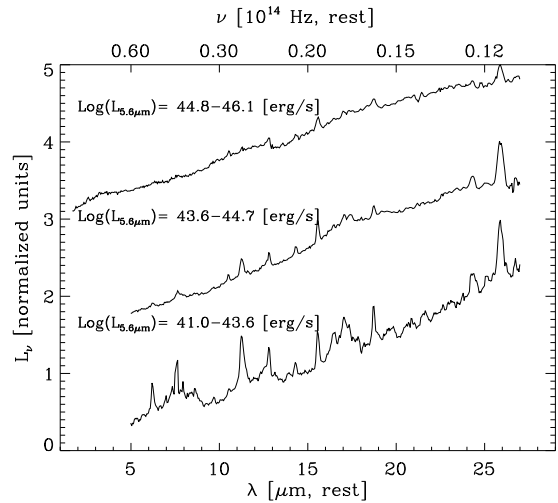}
	\caption{Rest-frame quasar composites of all three luminosity bins. Each template has been shifted in the ordinate for clarity, and the vertical location does not represent its true difference in $L_{\nu}$. The $\log(L_{5.6\micron})$ ranges of the objects that contributed to each bin are indicated above each template.}
	\label{fig:allcomps}
\end{figure}

The objects were ordered according to $\log(L_{5.6\micron\/})$, and
equally divided into three bins (with each bin containing 61 objects)
with the $\log(L_{5.6\micron\/})$ ranges for each tertile of
41.0--43.6, 43.6--44.7, and 44.8--46.1~[erg/s], respectively. For each
bin, a template spectrum was generated, as shown in
Figure~\ref{fig:allcomps} (all template spectra can be found in
Table~\ref{tab:template}). Template spectra were also constructed for
`reddened' and `normal' quasars according to their relative color,
$\Delta(g-i)$, (see Section~\ref{sec:sample}) with no significant
difference between them. We therefore conclude that the relative $g-i$
colour does not notably affect the MIR spectra.

\begin{figure}
	\centering
	\includegraphics[width=\linewidth]{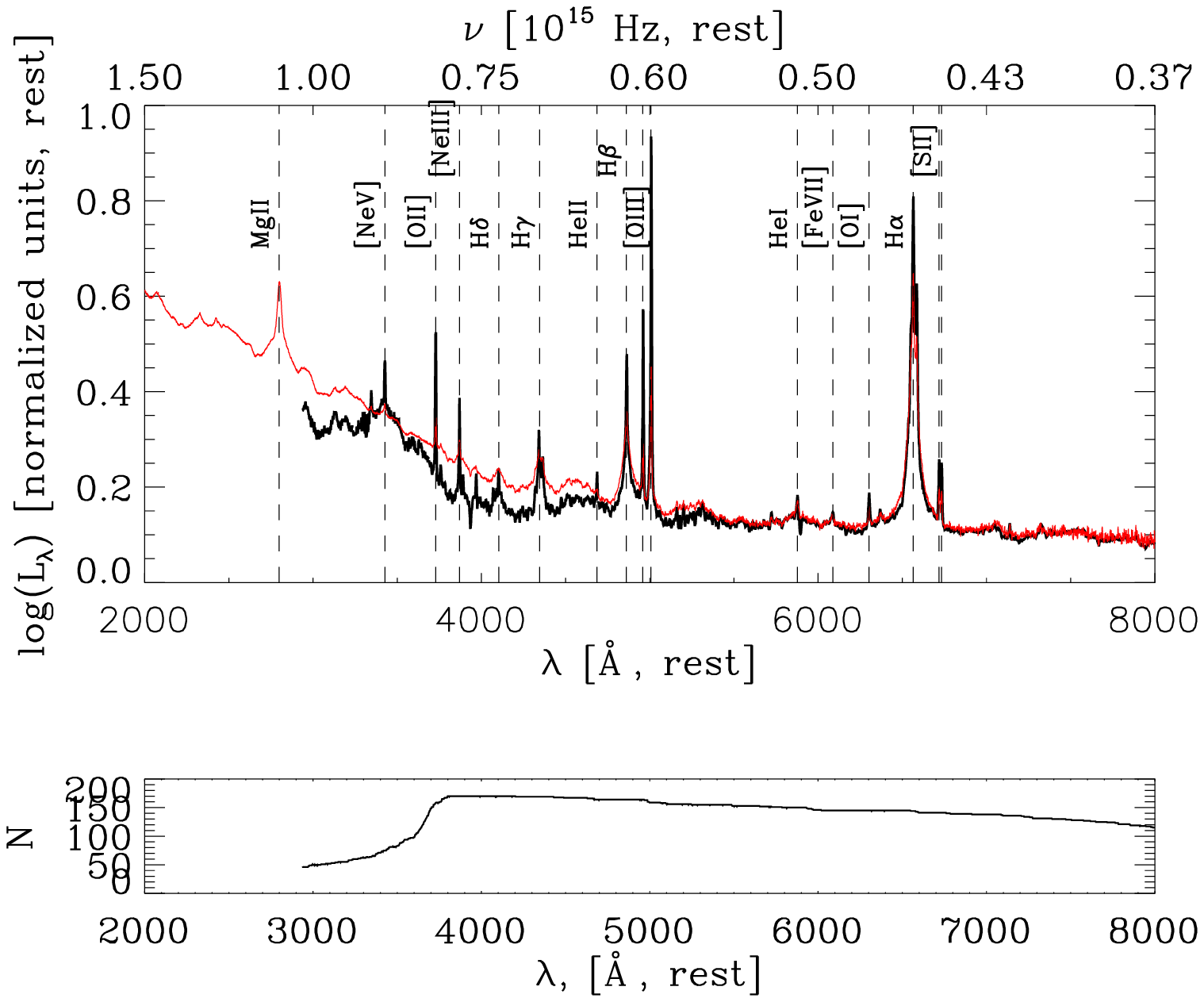}
	\caption{{\it Top panel}: Rest-frame UV-optical quasar template of all objects in the sample (solid black line). Overplotted is the \citealt{van01} composite (solid red line), normalized to the flux at $7481$~\AA. Prominent emission lines are labeled. {\it Bottom panel}: Number of objects contributing to the template in each wavelength bin. Our template is consistent with the \citet{van01} composite at long wavelengths, but deviates at shorter wavelengths. The \citet{van01} composite has a larger contribution from many higher redshift (and consequently higher luminosity) objects, particularly at short wavelengths. As more luminous quasars typically have bluer continua (e.g., \citealt{ric02}), the discrepancy between our composites shortward of $\sim4600$~\AA\ is not unexpected.}
	\label{fig:total_sdss}
\end{figure}

\begin{figure}
	\centering
	\includegraphics[width=\linewidth]{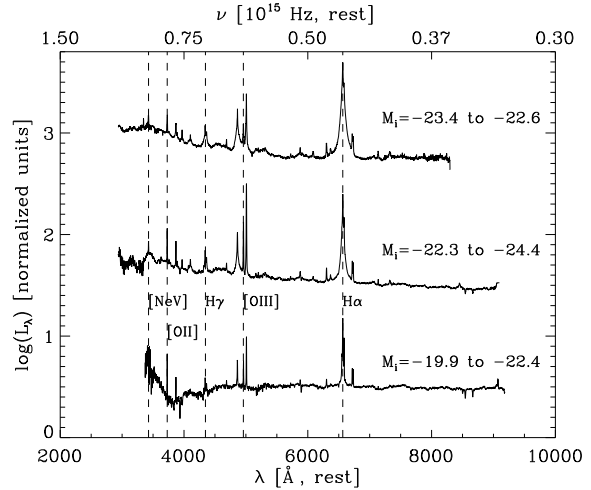}
	\caption{Rest-frame optical quasar templates. Each template has been shifted in the ordinate for clarity, and the vertical location does not reflect the true difference in $L_{\lambda}$. The uppermost template contains the same objects as the uppermost template of Figure~\ref{fig:allcomps}. The range in $M_{i}$ of the objects contributing to the template are labeled above each template.}
	\label{fig:allcompssdss}
\end{figure}

As previously mentioned, each of our objects with IRS spectra also
has a counterpart in the SDSS quasar catalogue. We provide SDSS
template spectra (Figure~\ref{fig:total_sdss}), which were generated
in the same manner as the IRS templates (i.e., using
$\sigma$~clipping), and binned according to their 5.6~$\micron$
  luminosity. For the SDSS templates, we use a bin size of
$2$~\AA. Corresponding luminosity-binned template spectra are also
provided (Figure~\ref{fig:allcompssdss}). The SDSS spectra cleaned of
OH sky lines were taken from the \citet{hew10} quasar catalogue.

\begin{table}
    \caption{Template Spectra for all objects, and luminosity binned objects (full table available in electronic format)}
    \centering
    \begin{tabular}{@{}lcccc}
      \hline\hline
       & \multicolumn{4}{c}{$\rm{L_{\nu}}$ [erg/s/Hz, normalized units]}\\
      $\lambda~[\micron]$ & $\rm{L_{\nu~tot}}$ & $\rm{L_{\nu~1}}$ & $\rm{L_{\nu~2}}$ & $\rm{L_{\nu~3}}$ \\
      \hline
      1.6909 & 0.1088 & 0.1088 &   &   \\
      1.7409 & 0.1040 & 0.1040 &   &   \\
      1.7909 & 0.1115 & 0.1115 &   &   \\
      1.8409 & 0.1299 & 0.1299 &   &   \\
      \multicolumn{5}{c}{:} \\
      \hline
    \end{tabular}
    \vspace{1mm}
    \begin{flushleft}
      $\rm{L_{\nu~tot}}$ is the template for all objects. $\rm{L_{\nu~1}}$, $\rm{L_{\nu~2}}$, $\rm{L_{\nu~3}}$ are the template spectra for the most luminous objects, the intermediate luminosity objects and the least luminous objects respectively. Blank values are due to a lack of wavelength coverage because of the dependence of the luminosity on redshift (coverage for other templates begins between $3$---$5~\micron$).
    \end{flushleft}
    \label{tab:template}
\end{table}

\subsection{MIR Spectral Fitting}

\begin{figure*}
	\centering
	\includegraphics[width=\linewidth]{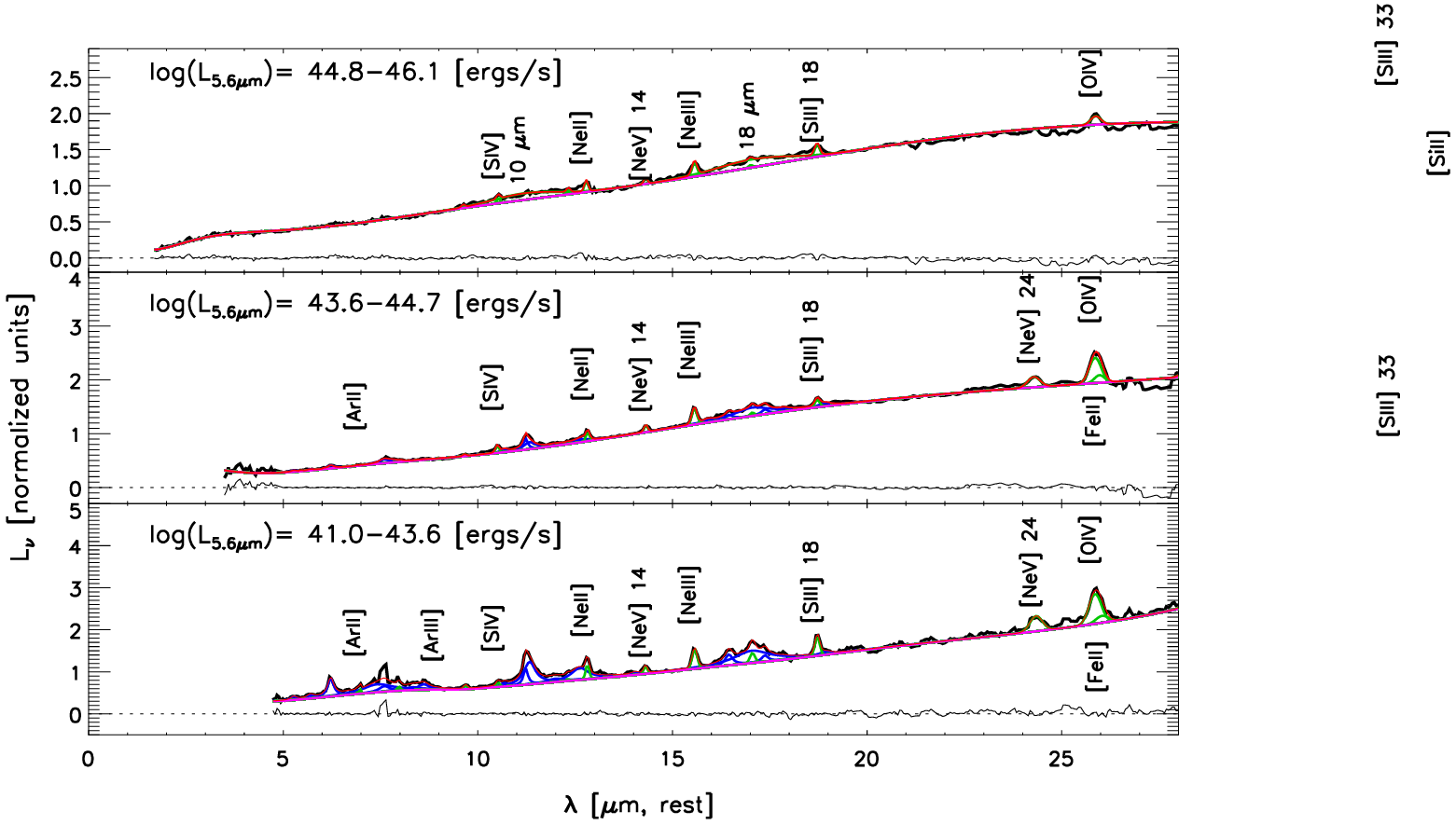}
	\caption{PAHFIT models of rest-frame templates, with residuals plotted (straddling the dotted black line at $L_{\nu}=0$). The top panel corresponds to the most luminous template, and the luminosity decreases towards the bottom panel. Green and blue curves show the narrow line and PAH features respectively. Magenta is the continuum fit, and red is the total model. Note the absence of PAH features in the top panel, and the that the 10 and 18~$\micron$ features only become visible in the top panel.}
	\label{fig:pahfit}
\end{figure*}

For each MIR template, we calculated line luminosities, central
wavelengths and equivalent widths for the detected features which
include narrow emission lines and PAHs. Because of the blending of
PAH and narrow line features in the composite, especially at shorter
wavelengths, we utilized the spectral decomposition code \verb$PAHFIT$
developed by \citet{smi07} to calculate the equivalent widths, central
wavelengths and relative intensities for MIR features. \verb$PAHFIT$
is an IDL script which uses least-squares minimization to de-convolve
spectra into their continuum, dust and narrow-line components (see
\citealt{smi07} for details). Because \verb$PAHFIT$ was not designed
with quasars and AGN in mind, it does not naturally fit warmer dust components
such as silicates in emission. To compensate for this, two broad,
gaussian functions were added to fit the 10 and 18~\micron\ silicate
features. The resulting fits for each composite appear in
Figure~\ref{fig:pahfit}.

Using the \verb$PAHFIT$-generated errors on the central intensity of
each, we estimated the error of the narrow line equivalent widths by
generating two gaussians (as the narrow line features are fit using
gaussians) with central peak intensities corresponding to the central
intensity $\pm$ the errors, and calculated the equivalent widths of
those gaussians. These two values are the estimated upper and lower
bounds for the \verb$PAHFIT$-generated equivalent widths.

\section{Results}

\subsection{Continuum Trends}

\begin{figure}
	\centering
	\includegraphics[width=\linewidth]{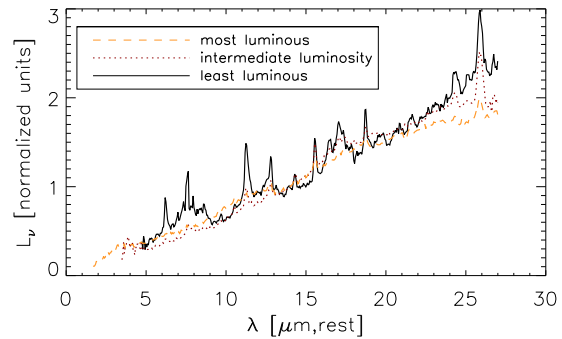}
	\caption{All luminosity binned templates plotted with no offsets (normalized to the template values at $14~\micron$. At longer wavelengths, the least luminous template (solid black line) rises to higher values than the intermediate luminosity template (dotted red line) and the most luminous template (dashed orange line).}
	\label{fig:blue}
\end{figure}

A comparison of the luminosity-binned templates of
Figures~\ref{fig:allcomps} and~\ref{fig:pahfit} show apparent trends
with increasing luminosity. The continuum becomes flatter redward of
$20~\micron$ (i.e., the higher luminosity templates are MIR-bluer than
the lower luminosity templates), which can be seen in
Figure~\ref{fig:blue}. The flattening of the continuum with luminosity
is gradual --- the most luminous template is MIR-bluer than the second
most luminous template, and the least luminous template is MIR-redder
than the second most luminous template. This continuum trend with
increasing luminosity is consistent with the findings of \citet{hao07}
who observed that the continuua of quasars were flatter than the
Seyferts in their sample.

The most luminous template was the only template to require the
inclusion of silicates in emission at both $10$ and $18~\micron$. Also
visible in the highest luminosity template is the $3~\micron$ (or NIR)
bump, which is likely caused by graphites located closer to the
central engine, as graphites have a higher sublimation temperature
($T_{sub}\sim1500$ K) than silicates \citep{bar87}. The presence of the
$3~\micron$ bump in higher luminosity quasars is consistent with the
anti-correlation between IR luminosity and the $1-8~\micron$ spectral
index noted by \citet{ric06b} and \citet{gal07}, and the requirement
of a hot dust component in the SED fitting of \citet{deo11} and
\citet{mor12}. However, we are able to detect the NIR $3~\micron$ bump
in the most luminous MIR template because our most luminous sources
tend to have higher redshifts (Figure~\ref{fig:diagnostic}); a
fraction of the rest-frame spectra have the advantage of covering this
range. Thus we cannot say for certain that this hot thermal dust
emission is absent in the lower-luminosity sources.

\subsection{PAH Features}

PAH features were detected in all but the most luminous template
(Figure~\ref{fig:pahfit}). If the most luminous template is supplied
to \verb$PAHFIT$ with no constraints on the PAH features, there will
be some PAH emission modeled in the $5-15~\micron$ region but almost
no PAH emission in the $15-20~\micron$ range. Since very few sources
exhibit this behavior \citep{smi07}, we interpret this as an artifact
of the software fitting the noise in the spectrum. We therefore
constrain the parameters of \verb$PAHFIT$ to not include PAH features
in the most luminous case, which is consistent with what is seen in
the spectral composite.

\begin{figure}

\begin{minipage}{\linewidth}
\centering
\subfloat[]{\label{noisy:a}\includegraphics[scale=.5]{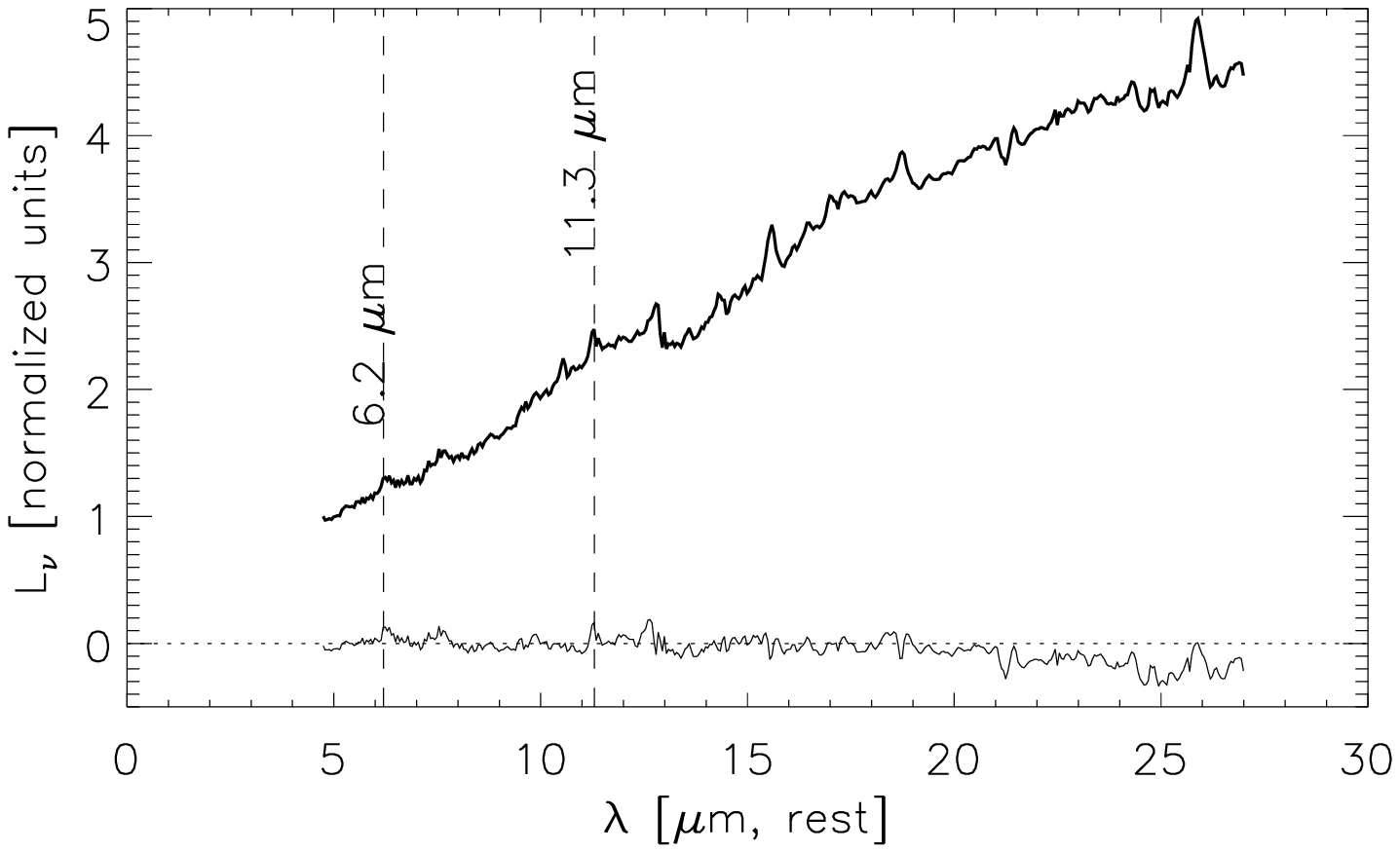}}
\end{minipage}\par\medskip
\begin{minipage}{\linewidth}
\centering
\subfloat[]{\label{noisy:b}\includegraphics[scale=.5]{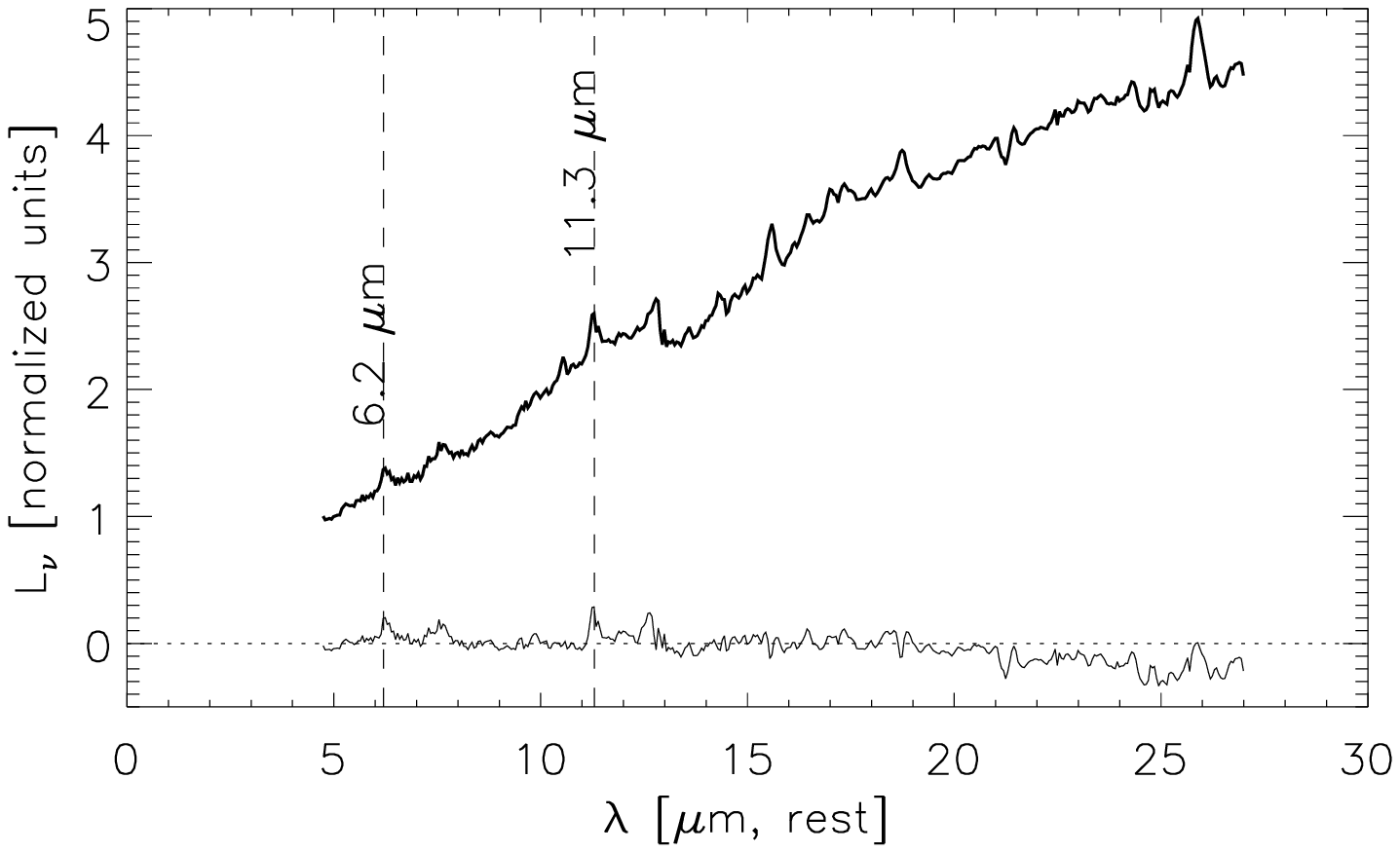}}
\end{minipage}\par\medskip

\caption{a) The model PAH features for the lowest luminosity template
  (with synthetic noise) added to the most luminous
  template. Residuals between the model fit of Fig.~\ref{fig:pahfit}
  and the PAH-added template straddle the dashed line. The PAH
  features are not detectable in the residuals above $2~\sigma$ (i.e.,
  they do not stand out significantly against the background noise).
  b) The PAH features from the lowest luminosity composite have been
  scaled up in luminosity by a factor of $2.5$. Here the $6.2$ and
  $11.3~\micron$ are detected at $2$ and $3~\sigma$ respectively.
  Given that the PAH emission is assumed to be powered by young stars,
  the host galaxies of the most luminous quasars could be actively
  star-forming. The lack of detected PAH emission in the most
    luminous quasar template does not preclude active star formation
    (which powers PAH emission) in their host galaxies at the same
    level observed in the less luminous quasar templates.}
\label{fig:noisy}
\end{figure}

For the templates which do show prominent PAH emission features, the
combined EW of these features decreases with increasing $5.6~\micron$
luminosity, until they either are overwhelmed by the continuum
emission, or are not present in the most luminous template. To
determine how luminous the PAH emission would need to be in order to
be seen above the noise we took the best-fitting PAH model shown
  in the bottom panel of Figure~\ref{fig:pahfit} and added gaussian
  noise ($S/N=14.5$, chosen from the S/N ratio at $6.2~\micron$).  To
  check the most optimistic case, the PAH model was assumed to come
  from the highest luminosity object in the lowest luminosity bin
  (($\rm \log(\rm L_{5.6~\micron})=43.5~\rm ergs~\rm s^{-1}$), and the
  composite was assigned the lowest luminosity ($\rm \log(\rm
  L_{5.6~\micron})=44.7~\rm ergs~\rm s^{-1}$) of the objects in the
  highest luminosity bin.  These scaled spectra were then added
  together to produce Figure~\ref{fig:noisy}.  Even in this best-case
scenario, there are no prominent PAH features visible in the residual
spectrum made by subtracting the best-fitting most luminous composite
model from the composite+PAH-feature spectrum (both are not detected
above $2~\sigma$).  When we increase the luminosity of the model PAH
features by a factor of 2.5 (Figure~\ref{fig:noisy}), the $6.2$ and
$11.3~\micron$ complexes become detectable at the $2$ and $3~\sigma$
level respectively. Assuming that PAH luminosity is correlated
  with the star formation rate \citep{sch06}, this suggests there
  could be a non-negligible amount of star formation activity `hiding'
  in the mid-IR composite of our most luminous quasar spectra.

\subsection{Narrow Lines}

\begin{figure}
  \centering
  \includegraphics[width=\linewidth]{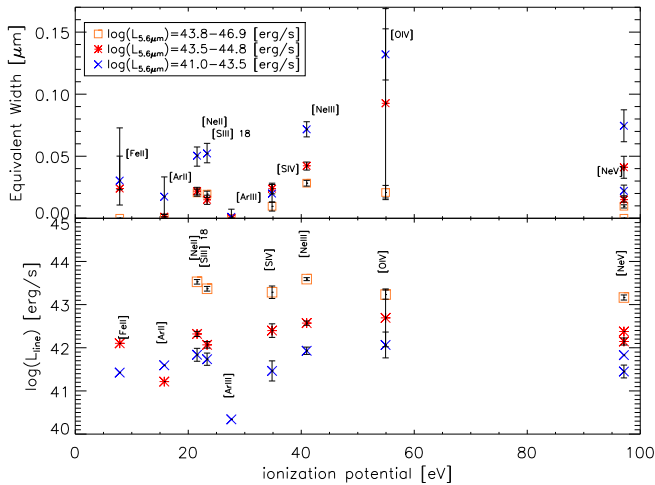}
  \caption{Equivalent widths (top) and line luminosities (bottom) of
    the narrow line features with their estimated errors plotted
    against the ionization potential. The data corresponding to the
    template containing the most luminous objects are marked in salmon
    squares. The next most luminous template is marked with red
    asterisks, with the least luminous template indicated by blue
    crosses. While the EWs of the narrow lines systematically decrease
    with increasing $L_{5.6\micron}$ luminosity, the luminosity in
    each line is still increasing with increasing continuum
    luminosity.}
  \label{fig:ew}
\end{figure}

From Figure~\ref{fig:ew} it is clear that the equivalent widths of the
narrow line and dust features are decreasing with luminosity,
consistent with \citet{hao07} (with the exception of [SIII]). The EW
is a measure of how the strength of the line compares to the
continuum, and is only physically meaningful if the continuum source
is the same as the ionizing source. In the case of the MIR spectra of
quasars, this is not the case. The narrow line region (NLR) is being
ionized by the UV through soft X-ray radiation from the accretion disk while the
MIR continuum is thermal emission from dust on much larger scales than
the accretion disk with temperatures on the order of $\sim100$s K. Considering this, there are three possible causes of the so-called MIR
Baldwin effect:

\begin{enumerate}

\item As the continuum strengthens, the lines themselves are weakening
  (i.e., the NLR contains less gas with increasing quasar
  luminosity).

\item The lines are maintaining the same strength (in terms of
  luminosity), but are being `washed out' in the strengthening
  continuum of higher luminosity objects (i.e., the NLR does not change
  with increasing quasar luminosity).

\item The lines are getting stronger, but are not keeping pace with the
  increasing power in the continuum (i.e., the NLR is scaling with
  increasing quasar luminosity, but the scaling is not linear).

\end{enumerate}

\citet{net04} suggested that the reason that narrow lines have
decreasing EW with increasing continuum luminosity is that the NLR
``disappears'' in some high luminosity quasars, and the effect of
averaging spectra with no narrow lines present to those with narrow
lines, artificially decreases the EW. This hypothesis is supported by
observed-frame near-IR observations of [OIII]$\lambda5007$~\AA\/,
using a luminous ($\log(L_{5100\AA})=45.77-47.38~[\rm ergs~\rm
s^{-1}]$), high-redshift sample of 29 quasars where they observed a
third of their sample to have either absent or very weak [OIII]
emission (see \citealt{net04} for further discussion).  They
speculated that the zone of [OIII] emission in luminous quasars could
be at distances beyond the extent of the host galaxy ISM. To test if
this effect is also seen in our lower luminosity (in comparison to
\citealt{net04}) sample, we looked at how the luminosity of
[NeV]~$14~\micron$ scaled with the $L_{5.6\micron\/}$ continuum
luminosity. [NeV]~$14~\micron$ is a good probe of the NLR because its
high ionization potential (of 97 eV) minimizes the potential
contribution from star formation in the host galaxy. Additionally, MIR
lines are less sensitive to extinction by dust compared to optical
lines. In order to obtain a luminosity from our normalized templates,
we needed an estimate of how the line strengths compare from template
to template. To do this, we integrated the \verb$PAHFIT$-generated
gaussians for each line, and multiplied that by the relative normalized luminosity at the appropriate continuum point for each
line, and then multiplied by the median $L_{5.6\micron\/}$ in that
luminosity bin (see Table~\ref{tab:linelum}). However, we find the
continuum is not well fit in the region around the [OIV] ($\sim 25 \mu
m$) line for the most luminous objects (see the top panel of
Figure~\ref{fig:pahfit}), and so the line luminosities in that
luminosity bin are likely underestimated. To account for this, we
measured the luminosity in the [OIV] line for the most luminous
composite by using a linear fit to the local continuum and fitting a
guassian to the line. We found the power in each line is still
increasing with luminosity (Figure~\ref{fig:ew}). Although the
continuum is outpacing the individual lines, the lines are still
getting stronger. This is consistent with the NLR becoming larger with
increasing continuum luminosity. However, \citet{net04} observed more
luminous objects than ours. We may just be constraining the limiting
luminosity, up to which the NLR does not ``disappear''.

\begin{table}
  \caption{MIR Spectral Feature Luminosity}
  \centering
  \begin{tabular}{@{}lccccc}
    \hline\hline
    & &  &High & Intermediate & Low \\
    & $\lambda^{a}$ & IP$^{b}$ & Luminosity & Luminosity & Luminosity\\
    Feature & $[\micron]$& [eV] & [erg/s] & [erg/s] & [erg/s] \\
    \hline
    H2 S(7) & 5.49 & & 42.97 & 41.76 & 40.51 \\
    H2 S(6) & 6.06 & & 43.07 & 41.50 & 40.69 \\
    H2 S(5) &  6.89 & & & 41.40 & 40.93 \\
    {[}ArII] & 6.97 & 15.76 & & 41.21 & 41.59 \\
    H2 S(4) & 7.98 & & & 41.36 & 41.69 \\
    {[}ArIII] & 8.97 & 27.63 & &  & 40.34 \\
    H2 S(3) & 10.52 & & 43.14 & 41.88 & 41.49 \\
    {[}SIV] & 12.31 & 34.79 & 43.28 & 42.39 & 41.46 \\
    H2 S(2) & 12.82 & & 43.11 & 41.05 & 40.86 \\
    {[}NeII] & 14.31 & 21.56 & 43.52 & 42.32 & 41.83 \\
    {[}NeV] 14 & 14.31 & 97.11 & 43.16 & 42.14 & 41.45 \\
    {[}NeIII] & 15.57 & 40.96 & 43.59 & 42.57 & 41.93 \\
    H2 S(1) & 17.06 & & 42.69 & 41.74 & 41.56 \\
    {[}SIII] 18 & 18.72 & 23.33 & 43.36 & 42.06 & 41.73 \\
    {[}NeV] 24 & 24.34 & 97.11 & & 42.37 & 41.82 \\
    {[}OIV] & 25.86 & 54.93 & 43.56$^{c}$ & 42.69 & 42.06 \\
    {[}FeII] & 26.04 & 7.87 & & 42.10 & 41.42 \\
\hline
\label{tab:linelum}
  \end{tabular}
  \vspace{1mm}
  \begin{flushleft}
    Line luminosities calculated for model outputs from PAHFIT for
    each luminosity template. Blank spaces indicate the feature was
    not detected by PAHFIT. \\$^{a}$Wavelengths are the best-fitting
    values from PAHFIT.\\ $^{b}$Ionization potentials taken from
    \citet{cox4}\\ $^{c}$Calculated from fitting the local continuum
    around [OIV] because the PAHFIT model was overpredicting the
    continuum at longer wavelengths. This value is used for
      Figures 12-14.
  \end{flushleft}
\end{table}

The bottom panel of Figure~\ref{fig:ew} demonstrates that the NLR is
growing with increasing continuum luminosity. We want to determine if
the scaling relation we find between the luminosity in various
forbidden lines and the 5.6~$\micron$ continuum luminosity is
consistent with what would be expected. First, we must determine what
the expected scaling relation should be, i.e., how does the number of
ionizing photons change with increasing 5.6~$\micron$ continuum
luminosity? For this calculation, we selected three forbidden lines
present in all three templates. We chose [NeV]~$14~\micron$ because of
its high ionization potential, $IP=97$ eV, which is unlikely to be
contaminated by star formation in the host galaxy. In addition, we use
[NeIII] ($IP=41$ eV) and [OIV] ($IP=55$ eV) as all three lines are
strong in all three composites, and cover a range of ionization
potentials.

\begin{figure}
  \centering
  \includegraphics[width=\linewidth]{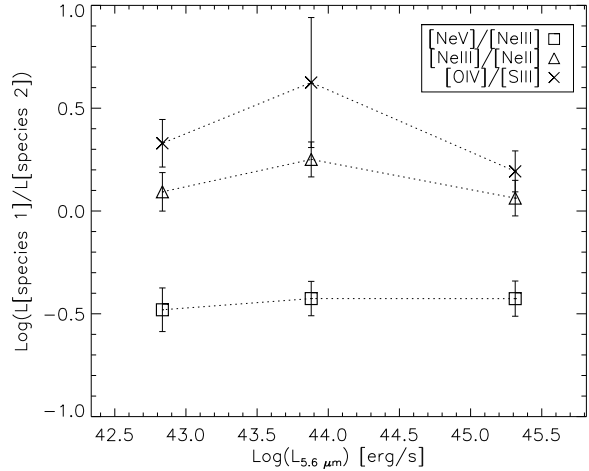}
  \caption{Line ratios of several MIR species ([NeV], [NeIII], [NeII], [OIV] and [SIII], plotted against the median 5.6~$\micron$ continuum luminosity of each template. The ratios involving Ne will be less sensitive to differences in abundances, and these are quite flat, indicating that the shape of the EUV does not change significantly with luminosity.}
  \label{fig:shape}
\end{figure}

To determine the number of ionizing photons for each species, we
integrated modified SEDs based on SEDs from \citet{kra13} and the
$\alpha_{\rm ox}$ relation (i.e., the anti-correlation between the
luminosity at $2500$~\AA\/ and 2~keV). We justify using the same SED
for each template because the shape of the SED does not vary
significantly from the optical to UV range as a function of
luminosity. Each SED is scaled to the median value of
  $L_{5.6~\micron}$ in each luminosity bin. All subsequent values used
  for integration are taken from these scaled SEDs. We used the
luminosity at $2500$~\AA\/ to estimate the 2~keV luminosity using
equation (5) of \citet{jus07}. We interpolate between the Lyman limit
and $2$~keV with a fixed power law. This assumption could be
inaccurate for calculating the ionizing flux, as the shape of the
extreme UV is not known, and may be a function of luminosity or
accretion rate (see discussion in section 6 of
\citealt{kra13}). However, we can constrain some behavior of the EUV
by examining narrow line flux ratios using species which probe
different ionization potentials. We measured how [NeV]/[NeIII],
[NeIII]/[NeII] and [OIV]/[SIII] change with $5.6~\micron$
luminosity. Figure~\ref{fig:shape} shows that the line ratios do not
change significantly across luminosity (within the
uncertainties). Therefore, we conclude that while we may be
underestimating the ionizing flux in all luminosity bins, there is no
evidence for a change in the continuum shape as a function of
luminosity between 41--97 eV.

\begin{figure}
  \centering
  \includegraphics[width=\linewidth]{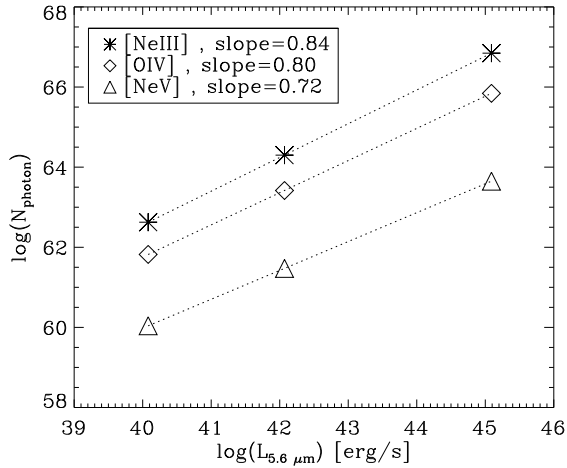}
  \caption{Number of ionizing photons integrated from a modified \citet{kra13} SED, plotted against the median 5.6~$\micron$ continuum luminosity of each template. The slopes for each linear fit are also provided on the plot. The number of ionizing photons increases more steeply with increasing continuum luminosity for the lower ionization species, as expected given the luminosity dependence of $\alpha_{\rm ox}$.}
  \label{fig:nphot}
\end{figure}

\begin{figure}
  \centering
  \includegraphics[width=\linewidth]{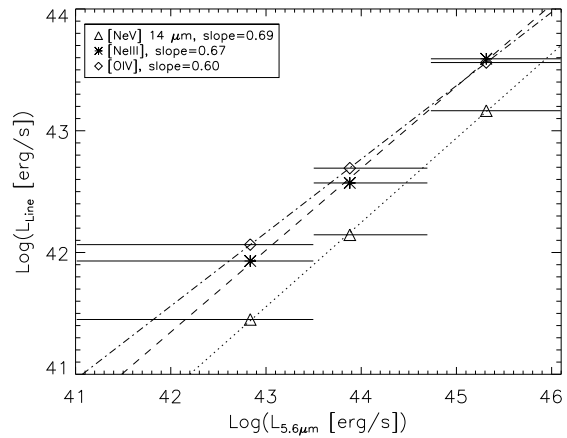} 
  \caption{[NeV], [NeIII] and [OIV] line luminosity plotted against
    the continuum luminosity at $5.6 \micron$. [NeV] is chosen because
    of its high ionization potential, limiting the contamination by
    star formation in the host galaxies. [NeIII] and [OIV] are chosen
    as they are strong lines present in all composites, and probe
    different ionization potentials. Horizontal bars indicate the
    luminosity range of the objects which contributed to that
    template. Dashed, dotted and dot-dashed lines are the linear fits
    to the points. The empirical slope for [NeV] of 0.69 is very close
    to the theoretical slope of 0.72 in Figure~\ref{fig:nphot}, and therefore we find no evidence for a ``disappearing'' NLR as probed by these lines.}
  \label{fig:disappear}
\end{figure}

With our modified SEDs in hand, we needed to determine appropriate
upper integration limits for each species. To do this, we calculated
the energy at which the cross-section for ionization would decrease to
half of its maximum (which occurs at the ionization energy). The
calculated ranges are $41$---$68$~eV, $55$---$102$ eV, and
$97$---$147$~eV for [NeIII], [OIV] and [NeV] respectively. The
relationship between the number of ionizing photons and the
5.6~$\micron$ continuum luminosity can be found in
Figure~\ref{fig:nphot}. We find that for lower ionization species, the
number of ionizing photons rises more steeply with increasing
5.6~$\micron$ luminosity, than for higher ionization species. This
expected scaling relation assumes the gas densities and ionization
parameters of the NLR gas are similar in all objects. All species had
predicted slopes within the range of $0.72$---$0.84$. We plotted the
line luminosities of [NeV], [OIV] and [NeIII] against the
5.6~$\micron$ continuum luminosity (Figure~\ref{fig:disappear}),
including linear fits to each species, with observed slopes ranging
between 0.60---0.69. These best-fitting values fall below the expected
range, but given the uncertainties, we consider them to be
consistent. In addition, [NeIII] may have a significant contribution
from star formation in the host, especially in the lower luminosity
objects with relatively more power coming out in X-rays; this would
have the effect of reducing the slope. [NeV] is not expected to have a
strong contribution from star formation, and we find the slope of
Figure~\ref{fig:disappear} to be very close to the predicted slope of
Figure~\ref{fig:nphot}. In the stratified picture of the NLR, the more
highly ionized gas is located closer to the ionizing source, and will
have a smaller volume filling factor than that of lower ionization
species. In the case of lower ionization species (with respect to
[NeV] with $IP=97$~eV) such as [OIV] ($IP=55$~eV), the ISM of the host
galaxy could be exhausted at the distance where the appropriate
ionization parameter gas should be located.  That is, as the
appropriate ionization zone moves farther out in more luminous
objects, it can extend beyond the region of the host galaxy with any
appreciable ISM \citep{net04}.  However, we do not see this effect.
Therefore, the NLR is scaling as expected in all three lines.

\section{Discussion}

The primary differences between the three templates can be attributed to host
galaxy contamination. The large fraction of low luminosity objects
makes it more likely that host galaxy contamination will be
significant in the low and intermediate luminosity templates; at low
luminosities, the quasar contributes a smaller fraction of the total
light and the host galaxy begins to dominate the spectrum. The PAH
emission is likely from star formation in the host galaxy, and is not
from the region excited by the accretion disk in the vicinity of
${R}_{\rm sub}$, as the hard ionizing radiation from the quasar would
destroy PAHs. The PAH features are most prominent when the MIR
continuum beyond $20~\micron$ is at its reddest in the least luminous
template, suggesting the presence of colder dust in the low luminosity
objects relative to the high luminosity objects. The continuum
emission at the red end of the MIR spectrum (beyond $20~\micron$) is
likely from cooler dust heated by star formation, as a turnover in
continuum emission is expected in the case of a pure AGN
\citep{deo09}. This is consistent with the results of \citet{sch06},
who studied a sample of 26 PG quasars, 11 of which show the
$7.7~\micron$ PAH feature. The 11 quasars which exhibited strong PAH
emission had enhanced FIR emission, and stronger emission from low
excitation lines, indicative of active star formation. They also find
the silicate emission feature at $10~\micron$ to be weaker in quasars
with detected PAH features; this seems to suggest that the star
formation is contributing a large fraction of the overall light,
making typical quasar features (e.g., silicates in emisson) less
prominent due to host galaxy contamination. This is consistent with
our understanding that at lower luminosities, the host galaxy is
contributing to a larger fraction of the total light, `washing' out
the quasar.

\begin{figure}
	\centering
	\includegraphics[width=\linewidth]{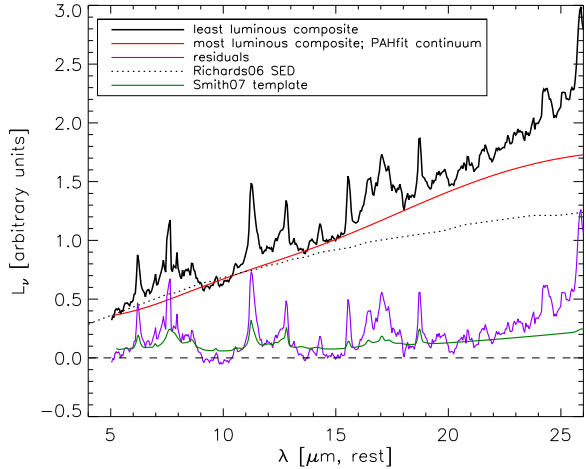}
	\caption{Comparison of the continuum of the most luminous
          template to the least luminous template.  The PAHFIT model
          continuum of the most luminous template (solid red line) was
          normalized to the least luminous template at $5~\micron$
          (solid black line). The \citet{ric06a} SED (dotted black
          line) is overplotted for comparison. The residuals from the
          subtraction of the PAHFIT model continuum (solid purple
          line) are above the 0.0 line (dashed black line). A
          starburst galaxy template from \citet{smi07} (solid green
          curve) is normalized to the $12.7~\micron$ complex in the
          residuals. The features in the residuals can largely be
          accounted for by star formation in the host galaxy; this
          suggests that the underlying quasar continuum is the same
          for all luminosity templates.}
	\label{fig:cont}
\end{figure}

To illustrate the host galaxy contamination in the least luminous
template, we normalized the \verb$PAHFIT$ model continuum from the
most luminous template to the $5~\micron$ intensity in the least
luminous template and subtracted them (Figure~\ref{fig:cont}). Under
the assumption that the continuum from the most luminous template is
more representative of the intrinsic quasar continuum, the residuals
should reflect the host galaxy contribution (specifically from star
formation in the host). We overplotted a galaxy starburst template
from \citet[][their template 3]{smi07}, and normalized the template to
the $12.7~\micron$ complex of the residuals. The \citet{smi07}
template does a satisfactory job of reproducing many of the main
features of the spectrum, and the overall shape. We would not expect
the starburst template to fit our residuals perfectly, as star-forming
galaxies do not have identical SEDs, but instead exhibit a range of
characteristics. Also included in Figure~\ref{fig:cont} is the
\citet{ric06b} quasar SED normalized to the continuum model from the
most luminous template. Our model continuum matches well to the
\citet{ric06b} at shorter wavelengths.

Considering the lack of PAH emission, together with an expected
turnover in continuum in the red end of the MIR \citep{mor12}, and the
results of Figure~\ref{fig:cont} we suspect the most luminous template
is representative of the true MIR spectrum of a quasar. The MIR quasar
spectrum is marked by the presence of coronal lines with a range of
ionization states, solid state silicate emission features, and a hot
dust dump in the near-IR (from species such as graphite).

\section{Summary and Conclusions}
\begin{enumerate}

\item We have generated high S/N MIR templates from 184 quasars using
  archival {\it Spitzer} IRS data to bring out common but faint
  features in typically low S/N MIR quasar spectra.  We also
  constructed three luminosity-binned templates by ordering, and
  dividing the sample according to the luminosity at $5.6~\micron$.

\item Because of the observed change in continuum, paired with the
  presence/absence of PAH features in the template, we hypothesize
  that the most luminous template is more representative of the
  intrinsic quasar spectrum. The other templates have much more host
  galaxy contamination due to the relatively low luminosity of their
  AGNs which thus contribute a smaller fraction of the combined MIR
  emission.

\item In the luminosity-binned templates, we note several trends with
  increasing luminosity: the EWs of the narrow line and PAH features
  decrease, the luminosity in each narrow line increases, and the
  continuum flattens beyond $20~\micron$. The decrease in EW of the
  narrow lines with increasing luminosity is consistent with similar
  findings of a MIR Baldwin Effect, however, the EW is not a
  meaningful metric to use in the MIR because the ionizing continuum
  is in the UV to EUV, not in the MIR. We therefore recommend the line
  luminosity as a more meaningful metric.

\item PAH features are not detected in the most luminous template. We
  interpret this to imply that quasars are not the primary heaters of
  PAHs, and the PAH emission is likely from star formation in the host
  galaxy and does not scale with the AGN luminosity.  More luminous
  starbursts are required to be detectable with a luminous AGN
  continuum.

\item Silicate features at $10$ and $18~\micron$ are seen in emission,
  but only in the most luminous template. A `bump' around $3~\micron$
  is present in the most luminous template, which is likely from
  graphites (or other species with a high sublimation temperature)
  located closer to the central engine which may or may not be unique
  to the high luminosity template given the lack of redshift coverage
  in the lower luminosity templates.

\item Using the luminosity of [NeV], a high ionization MIR line
  insensitive to extinction and star formation in the host galaxy, we
  probe the size of the NLR and find it to be scaling as expected with
  the continuum luminosity.

\end{enumerate}

\section{Acknowledgments}

We thank the Natural Science and Engineering Research Council of
Canada and the Ontario Early Research Award Program for supporting
this work. Allison. R. Hill was supported by an Ontario Graduate
Scholarship. G.T.R. acknowledges the generous support of a research
fellowship from the Alexander von Humboldt Foundation at the
Max-Planck-Institut f\"{u}r Astronomie and is grateful for the
hospitality of the Astronomisches Rechen-Institut.  We would like to
thank Aycha Tammour and Jan Cami for helpful discussions. Finally, we would like to thank the anonymous referee for helpful comments that improved the presentation of this paper.

\bibliographystyle{mn2e}
\bibliography{aasbib}

\end{document}